\documentclass[reprint,amsmath,amssymb,longbibliography,aps, prb]{revtex4-2}
\pdfoutput=1
\usepackage{hyperref}
\usepackage{url}
\usepackage{color}
\usepackage{acro}
\usepackage{chemmacros}

\newcommand{\beq}{\begin{eqnarray}}
\newcommand{\eeq}{\end{eqnarray}}

\usepackage{graphicx}

\begin{document}

% \title{Magnetic anisotropy from strain-induced dislocations in correlated electron systems}
\title{ Magnetic anisotropy from  linear defect structures in correlated electron systems}

\author{Mainak Pal$^{1*}$\thanks{email for correspondence: mainak.pal@ufl.edu}, Laetitia Bettmann$^2$, Andreas Kreisel$^2$, and P.J. Hirschfeld$^1$ }

\affiliation{$^1$Department of Physics, University of Florida, Gainesville, Florida, USA}
\affiliation{$^2$Institut f\" ur Theoretische Physik, Universit\"at Leipzig, D-04103 Leipzig, Germany}

\begin{abstract}
Correlated electron systems, particularly iron-based superconductors, are extremely sensitive to strain, which  inevitably occurs in the crystal growth process.  
Built-in strain of this type has been proposed as a possible explanation for experiments where nematic order has been observed at high temperatures corresponding to the nominally tetragonal phase of iron-based superconductors.  Strain is assumed to produce linear defect structures, e.g. dislocations, which are quite similar to O vacancy chainlets in the underdoped cuprate superconductor YBCO.   Here we investigate a simple microscopic model of  dislocations in the presence of electronic correlations, which create   defect  states that can drive magnetic anisotropy of this kind, if spin orbit interaction is present.    We estimate the contribution of these dislocations to magnetic anisotropy as detected by current torque magnetometry experiments  in both cuprates and Fe-based systems.
\end{abstract}

\maketitle
\section{Introduction} 
Defects in strongly correlated electron systems often behave quite differently from their weakly interacting counterparts\cite{4mousq}.  In particular, if a system is close to a phase transition, defects can tip the balance between two competing states.
For example, it is well known that impurities can create local magnetic states, or induce other types of local electronic order.  Similarly, it is well known that strain can tune the competition between two or more orders\cite{Cao2011}.
The use of strain as a tool to tune the electronic properties of correlated materials is rising quickly, particularly in materials where there is strong coupling between lattice and magnetic degrees of freedom, where the prospect control of magnetic properties and spin-polarized currents by strain is a long-standing goal. Applied strain has been used to study electronic nematic order, which influences transport currents, as well\cite{Boehmer2016,Kuo2016}.  

At the same time, interest in ways of studying built-in or internal strain, which also affects electronic properties,  has grown.  In this regard, local probes like STM have been particularly powerful.  Internal strain can occur in the crystal growth process, and pin local order more efficiently than point defects.  One particular instance where  strain has been invoked is the observation of nematic order, i.e. breaking of $C_4$ symmetry of the Fe-based superconductors at temperatures {\it above } the global tetragonal-orthorhombic transition temperature $T_s$\cite{Kasahara2012}.  The conclusions of these authors were based on torque magnetometry experiments on very small, possibly single-domain orthorhombic platelike crystals glued to sample holders.  Rather than proposing an explanation in terms of extrinsic strain physics, Kasahara et al. proposed a ``meta-nematic" transition based on a Ginzburg-Landau theory, where depending on the magnitude of a phenomenological coupling between lattice orthorhombicity and electronic nematic order, a $C_4$-breaking transition could occur at a higher temperature $T^*$, but yield only a very small symmetry breaking nematic field until the lower transition $T_s$ detected by x-rays.  To our knowledge, there is no microscopic theory justifying such a picture.  Similar signals of nematic order in the Fe-based superconductor NaFeAs were detected above the structural transition $T_s$, but in this case   these were indeed attributed to  strain\cite{Rosenthal2014}. 

In cuprates, a very similar situation exists in underdoped YBCO.  Strong indications of electronic nematicity have been reported in transport\cite{Ando2002} and low energy inelastic neutron scattering experiments\cite{Keimer2008}, in samples where the lattice orthorhombicity is extremely weak because the Cu-O chains are highly disordered.  The magnetic anisotropy measured by neutrons, which has a peak around O concentrations of about 6.4,   is nevertheless found to correlate strongly with the remanent $b$ direction of the chains, and is thought to be related to correlation-induced magnetism in the partially filled chains.  The influence of O vacancy ``chainlets", short vacancy segments in the chains that are inevitably formed in the doping process\cite{Veal1990}, was invoked in the transport work\cite{Ando2002}, and has been investigated theoretically in connection with the unusual low-temperature NMR lineshape\cite{Chen2009}.    

Recently, torque magnetometry was performed on YBCO,  showing that magnetic anisotropy was directly observable in experiment, and  increased sharply below  the pseudogap temperature\cite{Matsuda2017}.  Unfortunately, these 
measurements were only performed for O dopings where the chains should be fairly developed, above O6.5.  In this region it is believed that the trivial symmetry breaking in the electronic structure induced by the chains, together with spin-orbit coupling, controls the susceptibility anisotropy.  Indeed, the torque magnetometry signal is found to decrease as one underdopes.  

We expect that the O vacancy chainlets will induce 1D local magnetism and thereby control the nematicity observed at very low O concentrations.   These defects have in fact been imaged in STM\cite{delozanne1992},  and observed by NMR\cite{Yamani2006}.   As discussed in Refs. \onlinecite{Andersen2007,Chen2009,Schmid2010}, as one underdopes the correlations increase, thereby leading to enhanced magnetic effects.  These 1D defect structures are therefore a natural source of magnetic anisotropy.  

On the other hand, in neither the Fe-based nor the cuprate case has there been an attempt to understand how strain or chain-driven electronic anisotropy can couple to an external magnetic field as in, e.g. a torque magnetometry experiment, which measures magnetic susceptibility anisotropy, i.e. $\chi_{xx}-\chi_{yy}$.  Even in a system with  anisotropic ($C_2$ symmetric) electronic structure of whatever origin, local or global, {\it no torque will be produced unless the electronic anisotropy couples to the spin response, i.e. there must be significant spin-orbit coupling to produce a torque.  }

In other contexts,  highly anisotropic emergent linear defect states arising from pointlike potentials in strongly nematic superconductors  were studied in Ref. \onlinecite{Gastiasoro2014},    and line defects in unconventional superconductors have been suggested as generators of 1D topological superconductivity\cite{PhysRevX.11.011041}.  To our knowledge, torque magnetometry has not been applied in these cases, but it is a promising technique to learn about such systems.  

To  interpret torque magnetometry and other experiments, one needs a microscopic theory of how 1D-type defects created by built-in strain  or in the doping process couple to nematic or magnetic order, and thereby influence the system's coupling to external fields.    Such uniaxial defect structures are clearly visible in STM experiments\cite{Rosenthal2014}, but their magnetic character has not yet been probed.    Our predictions for induced magnetism in such cases can then be studied on systems with sufficiently smooth surfaces by spin-polarized STM. 

Calculating the local magnetic structure is not sufficient, however; one needs to understand how spin-orbit coupling allows an in-plane field to couple to strain.  Here we take a first step towards creating a microscopic theory of linear defects useful for studying the effect of strain on correlated electron systems and predicting the magnetic susceptibility anisotropy necessary to calculate the torque produced by an in-plane field.   We establish simple models of dislocations, with  lattice structure in their vicinity  relaxed by molecular dynamics, and calculate their magnetic character in the presence of Hubbard-like electronic correlations.  We then calculate the torque directly in the presence of the spin-orbit interaction.

To avoid the computational complications of multiorbital systems, we begin by studying a simple one-band model based on the cuprates.   Our hope is to make predictions for torque magnetometry experiments on the highly underdoped regime of YBCO, where torque magnetometry has not yet been attempted.   We do not, however, attempt to model  any particular realistic cuprate; rather, we make a simple model of a dislocation in a correlated system to see what kind of magnetic states can be created, as well as how they couple to an external field via the spin-orbit coupling in the crystalline environment.  Since torque magnetometry is evidently able to detect the nematic effect of the full chain at higher O concentrations in, we predict that it should be sensitive to the same enhanced  nematic effects driven by magnetism as detected by neutrons\cite{Keimer2008} and transport\cite{Ando2002}.

Mean field theory is used to study the effects of the local Coulomb interaction driving magnetism.  While such a method is known to overestimate magnetic order, it is easily adapted to inhomogeneous problems like the one at hand. We attempt then to estimate the magnitude of the torque  obtained with a crystal with a reasonable density of such dislocations, and discuss comparison to experiments on a variety of materials.

While our approach is fairly crude and aims to capture qualitative effects, it is to our knowledge, the only attempt until now to address how strain-induced dislocations can give rise to nematic behavior.

\section{Model}

\begin{figure}[tb]
    \centering
    \includegraphics[width=0.9\linewidth]{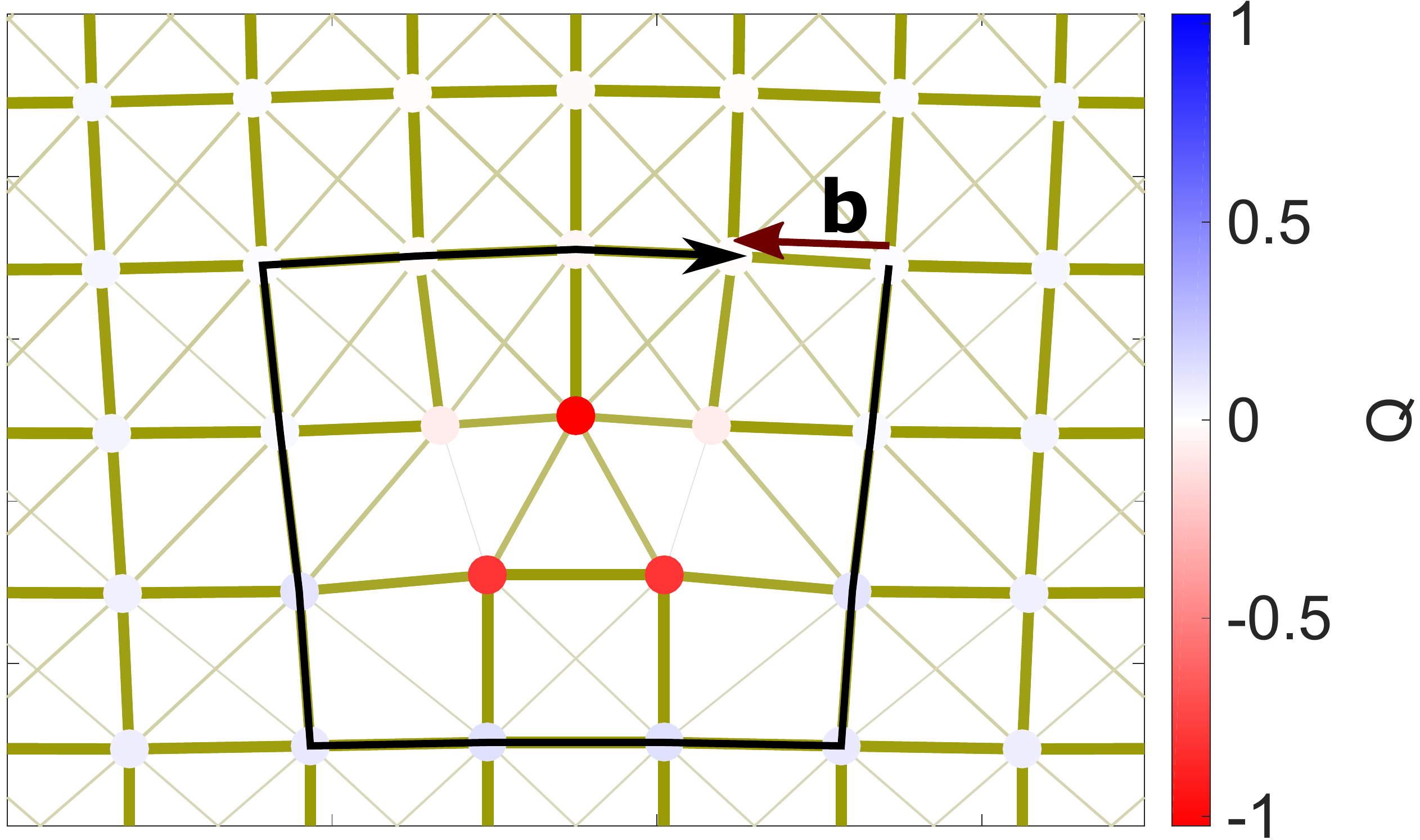}
    \caption{Schematics  of a single dislocation in a two dimensional lattice. Away from dislocation, sites are bonded to four nearest neighbors (NN), as in undistorted lattice. Interstitial space near the dislocation (not because of extra space, but suitable bond length) can however accommodate additional atoms mediating additional local bonds and an effective charge accumulation. Color of the dots represents effective charge accumulation calculated from valence bond sums [\onlinecite{Graser2010}]. Thicker bonds indicate larger magnitude of hopping. Non-zero Burgers $\mathrm{\textbf{b}}$ vector is shown by brown arrow. 
    }
    \label{fig:schematic_dislocation}
\end{figure}

In presence of strain, an otherwise homogeneous lattice can undergo dislocation in several ways.
Fig. \ref{fig:schematic_dislocation} shows a schematic diagram of a single edge-dislocation in two dimensional square lattice.    
In a homogeneous lattice, each site is connected to four nearest neighbors (NN), but dislocation may lead to space available for interstitial atoms, for example  excess oxygen atoms in case of cuprates, which lead to additional bonds and thereby a corresponding effective charge formation locally, as shown in the figure.   
To investigate the interplay of dislocations and effects of correlations that might induce local magnetism, we use an effective model Hamiltonian that describes one electronic state per unit cell and take two effects of the dislocation into account: a) the changes of the effective hopping for electrons moving between the lattice sites b) possible charge accumulation leading to an onsite potential.
\subsection{Dislocation pair}
Here, we consider almost covalent bonds between the  atoms in the non-defective system, such that missing atoms lead to a more ionic electronic configuration.  
In cuprates, for example, the Cu-O-Cu bridges mediate the hopping between sites, and excess oxygen atoms then lead to an effective charge.
These charge inhomogeneities, which can be calculated crudely by valence bond sums (see, e.g. Ref. \onlinecite{Graser2010}), together with locally  modified hopping amplitudes, can contribute to the nucleation of local magnetic or other orders.    Of course, the charge and the corresponding potential, as well as the hopping matrix elements, are  not only dependent on the number of bonds, but also on the distance between neighbors.
We include these charge inhomogeneities in the calculations by considering
them in the effective Hamiltonian as onsite potentials.  We will refer in what follows below to a cuprate model, and discuss prospects for observing these effects in cuprates, although the model is in fact very general and can describe similar phenomena in a variety of systems.  

A solitary edge-dislocation as shown schematically in Fig. \ref{fig:schematic_dislocation} has a nonzero Burgers vector  
for any loop surrounding the region of dislocation. But in a realistic macroscopic sample,  a second nearby dislocation often cancels the effect of the first, such that the effect of dislocation  diminishes rapidly as one moves away from the  defect. As a result it is useful and much easier computationally to study a dislocation-antidislocation pair, to ensure that the Burgers vector is zero.  Such pairs can occur in other configurations, such as the edge dislocations discussed in TiN by Yadav et al.\cite{Yadav2014}, which we do not discuss in this proof of principle work.

\begin{figure}[tb]
    \centering
\includegraphics[width=1.03\linewidth]{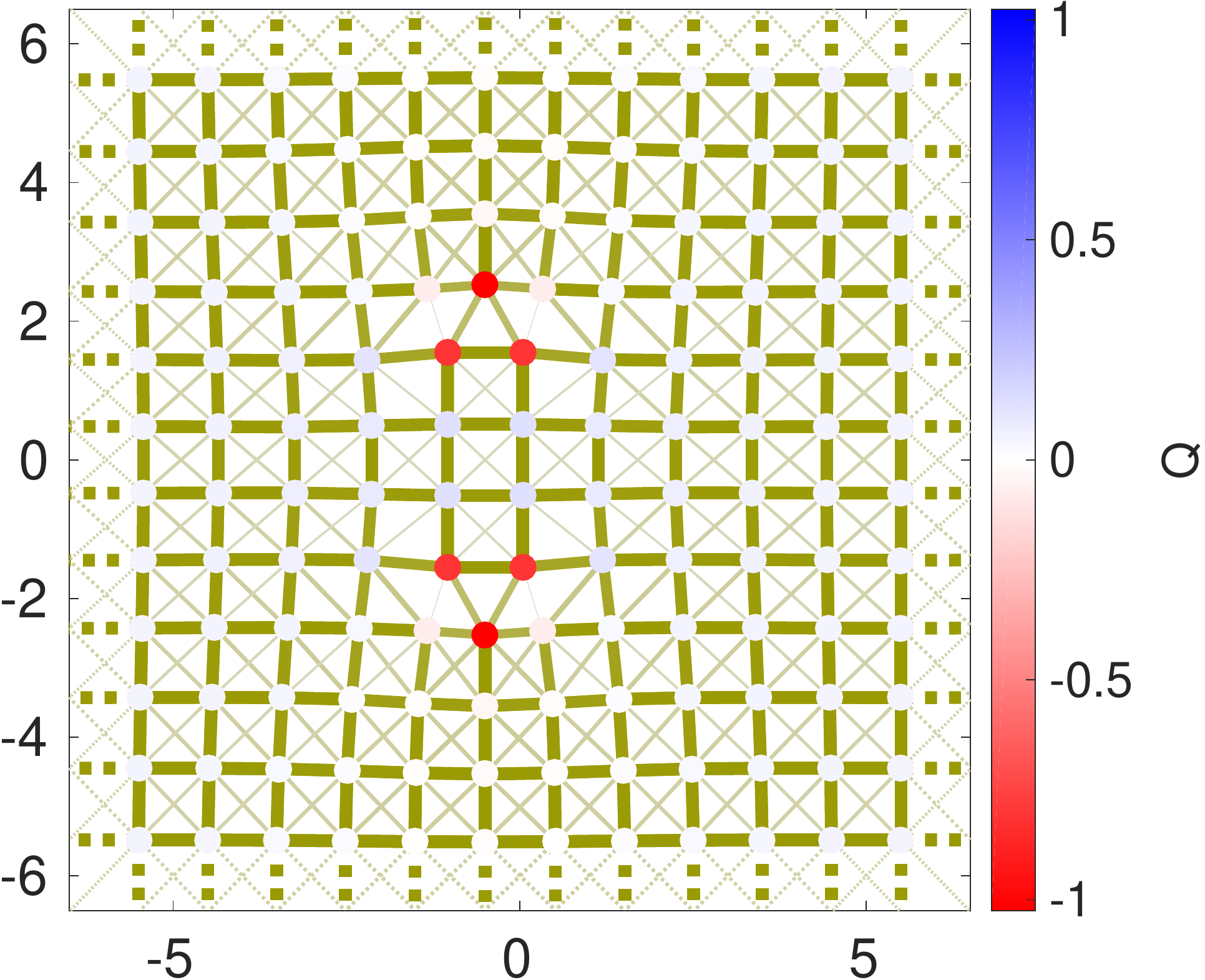}
    \caption{Distortion and charge formation 
    for a $4$ site long  dislocation-antidislocation pair in a $12\times12$ lattice.  {Thickness of the bonds parameterize the strength of the intralayer hopping matrix elements $t(\textbf{r}_{ij})$, while the color of the dots (shown in the colorbar) at sites represents the effective charge value in units of the electron charge at sites  as determined by valence bond sums.   Only hoppings within a range of $|\textbf{r}_{ij}|=1.5$ are shown.  Dashed hoppings at sample border are periodically continued to the other side of the lattice.    } }
    \label{fig:Distortion_and_charge_formation_in_lattice}
\end{figure}

Fig. \ref{fig:Distortion_and_charge_formation_in_lattice} shows such a dislocation-antidislocation pair.
  The lattice site positions $\mathbf r_{i}$ as a result of dislocation and subsequent distortion  were simulated 
with Large-scale Atomic/Molecular Massively Parallel Simulator (LAMMPS)\cite{Plimpton1995}. For this, sites in the lattice were assumed to be bonded to the nearest and next nearest neighbors via unstretched springs of  the same strength in absence of any dislocations. Then lattice sites were selectively removed and the system was allowed to relax under  a harmonic potential of the stretched springs with periodic boundary conditions using a nonlinear conjugate gradient method \cite{Polak1969} (Fig. \ref{fig:Distortion_and_charge_formation_in_lattice}).
  
This approach gives a reasonable configuration close to the dislocation but is simpler than usual approaches with pair potentials. Note that the final result does not depend on the initial positions of the lattice points.  Irrespective of whether the boundary is kept fixed (which would mean external strain) or allowed to relax, the effect of interest, i.e. appearance of local magnetism as described in later sections, remains in the vicinity of the dislocation.
 The thickness and color of the bonds each independently represents the magnitude of the hopping between sites, such that a thicker and greener bond represents a stronger hopping. Dotted bonds (with the same thickness scheme and color scheme as solid bonds) represent hopping at the periodic boundary. 
 The color of the dots represent the effective accumulated charge due to interstitial atoms calculated via valence bond sums. Sites that are separated by a distance $|\textbf{r}_{i}-\textbf{r}_{j}|=|\textbf{r}_{ij}|<1.3$ (where $\textbf{r}_{i}$ is the two dimensional position of the $i$th site)   are accounted for in valence bond sums, and hoppings of range more than 2.05 are truncated. The precise protocols adopted for the construction of these sums and the hoppings $t$ (in plane), $t^\perp(\textbf{r}_{ij})$ (out of plane) in the dislocated system are discussed below.

 Fig. \ref{fig:Distortion_and_charge_formation_in_lattice} clearly shows how some hoppings are removed in the vicinity of the dislocated sites due to the increase in nearest neighbor distances induced by distortion, and how additional bonds are mediated by interstitial atoms, again accommodated by distortion, that contribute to the valence bond sums.     

\subsection{Homogeneous system: cuprate bilayer} 
\begin{figure}[tb]
    \centering
    \includegraphics[width=1\linewidth]{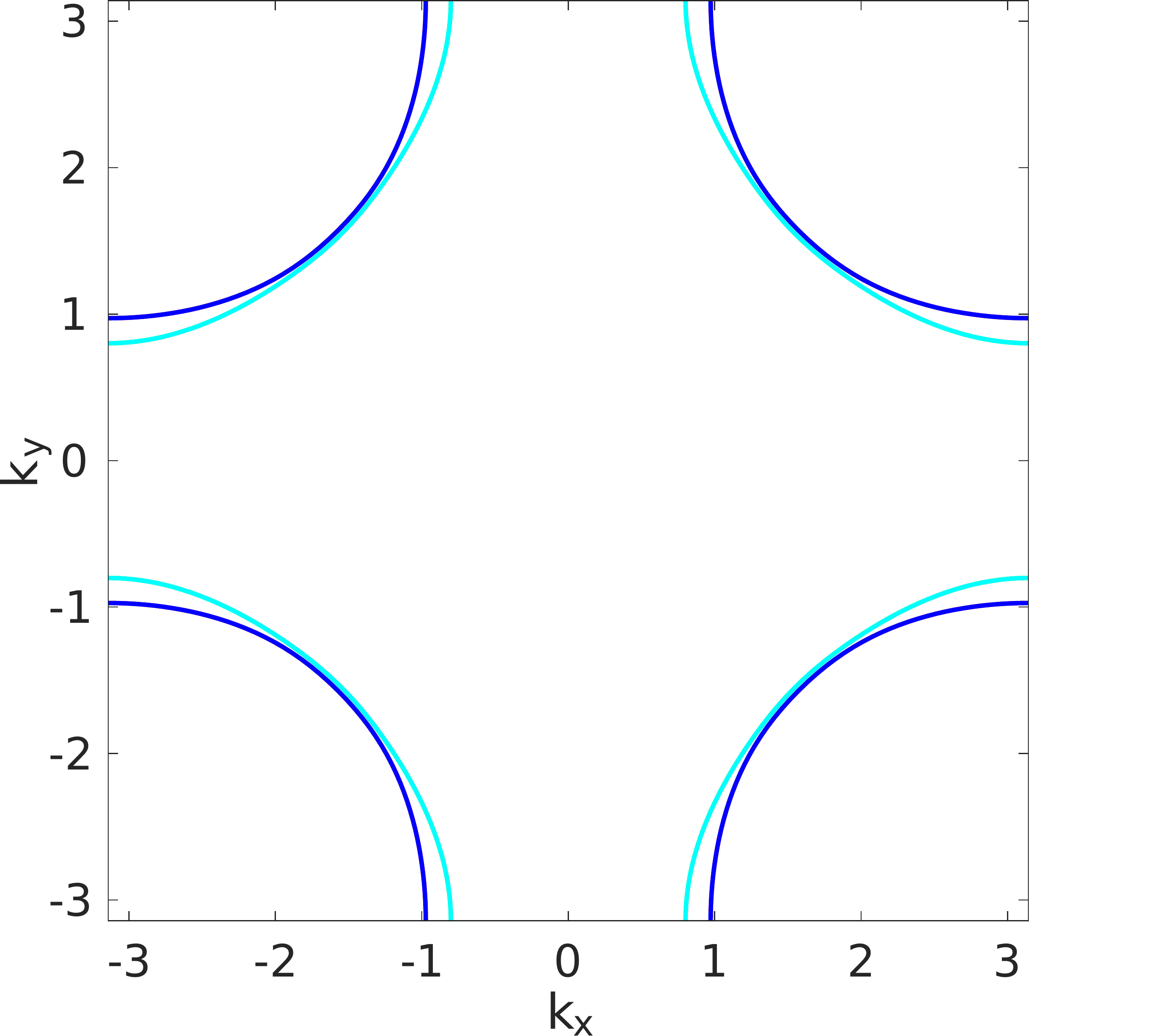}
    \caption{Fermi surface of the homogeneous bilayer system in presence of spin-orbit coupling. The bilayer hybridization splits each of the four fermi surfaces into two in a momentum dependent way keeping the four-fold degeneracy of the electronic states intact at four points in the $(|k_x|,|k_y|)=(\pi,\pi)$ directions. Spin-orbit term increases the splitting and lifts the degeneracy from four-fold to two-fold at the $(|k_x|,|k_y|)=(\pi,\pi)$ directions. 
    }
    \label{fig:fermi_surface}
\end{figure}

It is convenient to consider a bilayer rather than a single layer system, because it is significantly simpler to introduce the lattice version of spin-orbit coupling required to create anisotropy in the magnetic response, and thus to make predictions for torque magnetometry experiements.  In addition, we may then make contact specifically with experiments  performed on bilayer cuprates such as YBCO-123 and Bi-2212.
The Hamiltonian  for a homogeneous bilayer system includes intralayer  hoppings,
\begin{equation}
H_{\mathrm tb}  ={\displaystyle \sum_{
|\mathbf{r}_{ij}|\leq2
,\sigma}\left(t(\mathbf{r}_{ij})-\mu\delta_{i,j}\right)c_{i,\sigma}^{\dagger}c_{j,\sigma}}\,,
\label{eq:Htb}
\end{equation}
where $i,j$ are in a single layer, and we have taken non-zero hopping up to next next nearest neighbors (NNNN), i.e. the in-plane vector  $|\mathbf{r}_{ij}|=1,\sqrt{2},2$,  within the same layer. 
The NN, NNN and NNNN intralayer hoppings for the homogeneous lattice are respectively $t=-0.15\ \mathrm{eV}$, $t'=0.044\ \mathrm{eV}$, $t''=-0.002\ \mathrm{eV}$.   The interlayer hopping terms are given by
\begin{equation}
H^{\perp}_{\mathrm{tb}}  ={\displaystyle \sum_{
% \langle i',j'\rangle
|\mathbf{r}_{ij}|\leq 2
,\sigma}t^{\perp}(\textbf{r}_{ij})c_{i,\sigma}^{\dagger}c_{j,\sigma}}\,.
\label{eq:Htb_perp}
\end{equation}
where now the amplitude $t^\perp$ includes the hopping from the NN of $i$ in the other plane, and we include terms hopping to 
  NN, NNN, and NNNN sites, with corresponding values $t^\perp(1,0,0.5,0.25)$ with   $t^\perp=0.012\, \mathrm{eV}$. In Fourier space these terms give the usual interlayer dispersion $\propto (\cos k_x - \cos k_y)^2$ \cite{Gotlieb1271}.   These choices describe  a  cuprate-like  Fermi surface with small bilayer splitting of the two bands, as shown in Fig. \ref{fig:fermi_surface}. 
  
The spin-orbit interaction in the bilayer\cite{Gotlieb1271} 
\begin{equation}\label{eq:homo_SO}
\begin{split}
 H_{\mathrm SO} &=  (-1)^\nu \gamma \Big\{{\displaystyle \sum_{\langle i,j_{\mathrm y}\rangle}i \left( c_{i,\uparrow}^{\dagger}c_{j_{\mathrm y},\downarrow}-c_{j_{\mathrm y},\uparrow}^{\dagger}c_{i,\downarrow}\right)}\\
 &\ \ \ + {\displaystyle \sum_{\langle i,j_{\mathrm x}\rangle}\left( c_{i,\uparrow}^{\dagger}c_{j_{\mathrm x},\downarrow}- c_{j_{\mathrm x},\uparrow}^{\dagger}c_{i,\downarrow}\right)} \Big\}+\mathrm h.c.
\end{split}
\end{equation}
is also included in the homogeneous system, 
with suitably chosen $\gamma$ (here taken to be $4.5\ \mathrm{meV}$ in our calculation), where $\langle i,j_{\mathrm x}/j_{\mathrm y}\rangle$ represents intralayer nearest neighbors of $i$th site
along +x or +y direction and $\nu=0,1$ is the layer index.
When transformed into a momenum space representation, this spin-orbit term can be interpreted as an in-plane momentum dependent magnetic field with components $h_{x,\mathrm{eff}}= 2(-1)^\nu\gamma \sin(k_y)$  and $h_{y,\mathrm{eff}}= 2(-1)^\nu\gamma  \sin(k_x)$  which flips sign from one layer to the other. 

\begin{figure*}[bt]
\includegraphics[width=\linewidth]{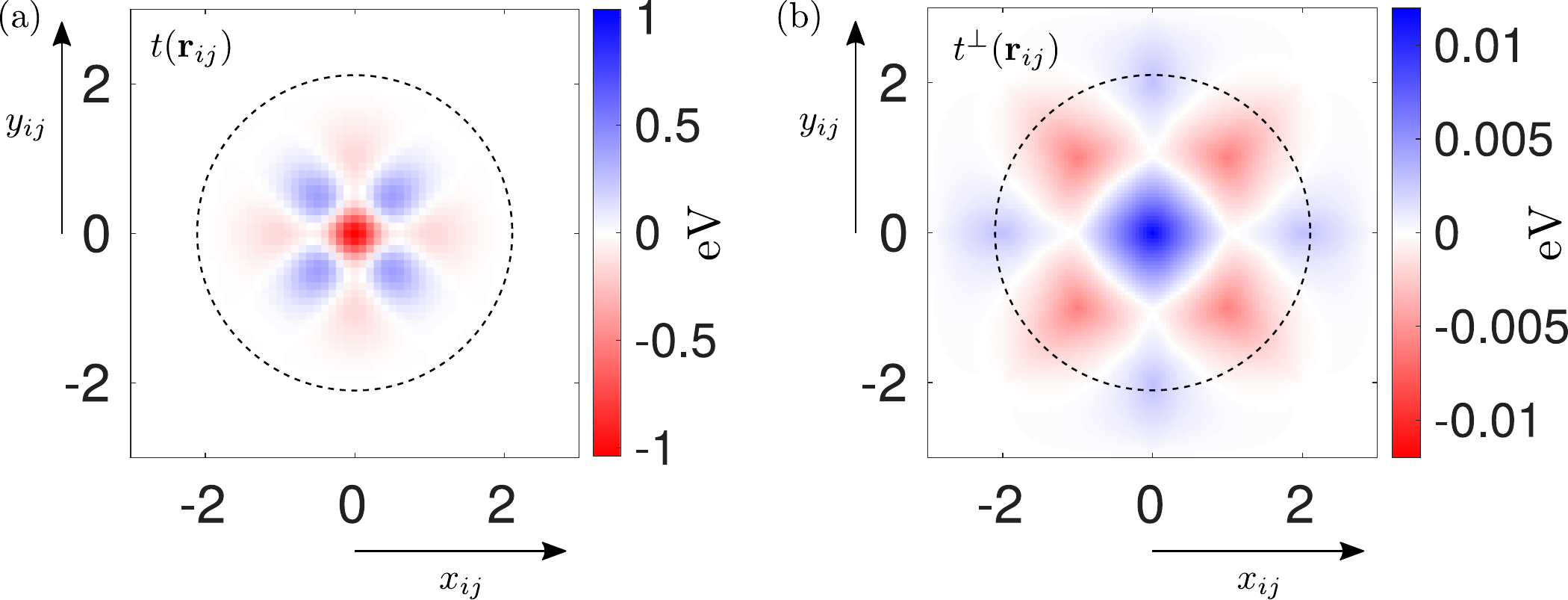}
\caption{Map, as discussed in model section, of (a) intralayer hopping integrals $t(\mathbf r_{ij})$  as function of the distance $\mathbf r_{ij}=(x_{ij},y_{ij})$ calculated from overlap of atomic $d_{x^2-y^2}$ orbitals at sites and then re-normalized to match cuprate $t$, $t'$, $t''$ model \cite{Norman1995} (b) interlayer hopping integrals $t^{\perp}(\textbf{r}_{ij})$, as function of the distance $\mathbf r_{ij}=(x_{ij},y_{ij})$, interpolated from what would be $t^{\perp}(\textbf{r}_{ij})$ in undistorted lattice \cite{Gotlieb1271}.   The circles represent range of hoppings retained in numerical calculations.}
\label{fig:intra_and_inter_layer_hopping}
\end{figure*} 
\subsection{Electronic structure of dislocation}
In a lattice distorted because of dislocation, the distinction among NN, NNN, NNNN is blurred and one needs to obtain both intralayer and interlayer hoppings as continuous functions of in-plane vector distance between sites $\mathbf r_{ij}=\mathbf r_{i}-\mathbf r_{j}$, i.e. as $t(\mathbf{r}_{ij})$  and  $t^{\perp}(\mathbf{r}_{ij})$. These maps are shown in Fig. \ref{fig:intra_and_inter_layer_hopping}.
The map for $t(\mathbf{r}_{ij})$ has been generated by computing the expectation value of the kinetic energy $-\nabla^2/(2m^*)$ for overlapping atomic $d_{x^2-y^2}$ orbitals, and the effective mass $m^*$ adjusted such that the NN and NNN hoppings for the homogeneous lattice agree roughly with those found in cuprate materials\cite{Norman1995}. { While this approach does not capture all details of { wave functions on}  neighboring atoms, 
%in the elementary cell, 
{ unlike e.g. the Slater-Koster method}, it does generate hopping elements as function of $\mathbf r_{ij}$ with the right symmetry and angle dependence.}
Both intralayer and interlayer hoppings in the distorted lattice are truncated beyond  $|\textbf{r}_{ij}|=2.05$ 
for the numerical calculations presented here  since these are small in magnitude beyond the the dashed circles in Fig. \ref{fig:intra_and_inter_layer_hopping}.

The spin-orbit term also undergoes modification because new NN bonds are formed which were neither  `x-neighbors' or `y-neighbors' in the undistorted lattice. A neighbor in the distorted system is considered `x-neighbor' or `y-neighbor' if it is closer to the `x-axis' or the `y-axis' respectively. So, the spin-orbit part of the Hamiltonian in the distorted system looks like
\begin{equation}
\begin{split}
H_{\mathrm{SO}} &= {\displaystyle \sum_{|\textbf{r}_{ij}|<1.3
,\sigma}(-1)^{\nu}\gamma_{ij}c_{i,\sigma}^{\dagger}c_{j,\bar{\sigma}}}\,,
\end{split}
\end{equation}
where $|\gamma_{ij}|=\gamma$ with sign decided by the proximity to x-axis or y-axis, as just described, and the spin index, by analogy to Eq. \ref{eq:homo_SO}. 

Finally,  dislocation  contributes an effective onsite potential in the Hamiltonian:
\begin{equation}\label{eq:disloc}
\begin{split}
H_{\mathrm Q} & ={\displaystyle \sum_{i,\sigma}V_{\mathrm{eff}}\ Q_{i}c_{i,\sigma}^{\dagger}c_{i,\sigma}}
 \end{split}
\end{equation}
An effective potential $V_{\mathrm{eff}}=1.5\ \mathrm{eV}$ was used in our calculation. 
The effective charge $Q_i$ at $i$th site, accumulating due to change in neighborhood and neighborhood distances $\textbf{r}_{ij}$ caused by dislocation, was calculated via valence bond sums \cite{Graser2010}
\begin{equation}\label{eq:charge_eq}
 Q_{i}=Q_{\mathrm Cu}+A\sum_{|\textbf{r}_{ij}|<1.3} Q_{\mathrm O}\exp(-\vert \textbf{r}_{ij}\vert ^2/\lambda^2)\,,
\end{equation}
where $Q_{\mathrm Cu}$,\ $Q_{\mathrm O}$ are copper and oxygen charges at $i$th site in case of cuprates. $Q_{\mathrm O}$ was taken to be $-2$ and $Q_{\mathrm Cu}$  was taken to be $+4$, representing effective charge of the copper ion along with the charge reservoir layers.
$A$ and $\lambda$ are two constants adjusted in such a way that $Q_i$ vanishes in undistorted lattice.

Thus the Hamiltonian used to study the dislocation is identical in form to Eqs. (\ref{eq:Htb}-\ref{eq:homo_SO}) except for the distorted lattice positions $\{\textbf{r}_i\}$, the ranges of the hoppings allowed, and the sign modifications of a small number of short bonds in the spin-orbit coupling.   In addition, it contains onsite potential terms due to charge transfer effects.  We have verified that results do not depend sensitively on small changes in the hopping and charging truncation ranges. 

\subsection{Electronic correlations}
As discussed in the introduction, defects can play an extraordinary role in correlated electron systems in proximity to competing ordered phases.  We consider here the situation, explored extensively in the cuprates\cite{4mousq,Andersen2007,Schmid2010}, where defects can freeze antiferromagnetic spin fluctuations locally in magnetic islands or other structures.  
We account for the short-range Coulomb repulsion with a  spin-rotationally invariant mean-field decoupling $H_\mathrm{U}$ of the usual Hubbard interaction
\begin{equation}
\begin{split}
H_{\mathrm U} & = U\,{\displaystyle \sum_{i}}\Big\{\langle c_{i,\uparrow}^{\dagger}c_{i,\uparrow}\rangle c_{i,\downarrow}^{\dagger}c_{i,\downarrow}+\langle c_{i,\downarrow}^{\dagger}c_{i,\downarrow}\rangle c_{i,\uparrow}^{\dagger}c_{i,\uparrow}\\
&\ \ \ -\langle c_{i,\downarrow}^{\dagger}c_{i,\downarrow}\rangle\langle c_{i,\uparrow}^{\dagger}c_{i,\uparrow}\rangle-\langle c_{i,\uparrow}^{\dagger}c_{i,\downarrow}\rangle c_{i,\downarrow}^{\dagger}c_{i,\uparrow}\\
&\ \ \ -\langle c_{i,\downarrow}^{\dagger}c_{i,\uparrow}\rangle c_{i,\uparrow}^{\dagger}c_{i,\downarrow}+\langle c_{i,\downarrow}^{\dagger}c_{i,\uparrow}\rangle\langle c_{i,\uparrow}^{\dagger}c_{i,\downarrow}\rangle\Big\}.
\end{split}
\end{equation}
One should note that the constant mean field terms, which are sometimes discarded for studying the dynamics, are important here to compare the energy of various states.

Finally, placing the system into  an in-plane magnetic field $\mathbf B$ will induce
a Zeeman term in the Hamiltonian
\begin{equation}
\begin{split}
H_{\mathrm B} & = - {\displaystyle \sum_{i}}\Big\{ h_{\mathrm x}\left(c_{i,\uparrow}^{\dagger}c_{i,\downarrow}+c_{i,\downarrow}^{\dagger}c_{i,\uparrow}\right)\\
 &\ \ \ +h_{\mathrm y}\left(-i\right)\left(c_{i,\uparrow}^{\dagger}c_{i,\downarrow}-c_{i,\downarrow}^{\dagger}c_{i,\uparrow}\right)\\
 &\ \ \ +h_{\mathrm z}\left(c_{i,\uparrow}^{\dagger}c_{i,\uparrow}-c_{i,\downarrow}^{\dagger}c_{i,\downarrow}\right)\Big\}\,,
\end{split}
\end{equation}
where $\mathbf h=g \mu_{\mathrm B}\mathbf B$ with the gyromagnetic ratio $g$ and the Bohr magneton $\mu_{\mathrm B}$. Thus, the full Hamiltonian that we consider is 
\begin{equation}\label{eq1}
\begin{split}
H & =H_{\mathrm{tb}}+H^{\perp}_{\mathrm{tb}}+H_{\mathrm Q}+H_{\mathrm{SO}}+H_{\mathrm U}+H_{\mathrm B}\,.
\end{split}
\end{equation}

\section{Results}
\subsection{Phase diagram of homogeneous system}
As discussed above, the intralayer and interlayer hoppings, together with the spin-orbit interaction, produce in the homogeneous system the standard low-energy electronic structure of a bilayer cuprate (Fig. \ref{fig:fermi_surface}).  The interlayer hoppings induce a momentum dependent splitting (the violet and cyan lines) of the Fermi surface of the monolayer system maintaining the four-fold degeneracy at the four nodal points in the $(|k_x|,|k_y|)=(\pi,\pi)$ directions. Further introduction of the spin-orbit term increases the splitting overall and lifts the degeneracy from four-fold to two-fold at the $(|k_x|,|k_y|)=(\pi,\pi)$ directions.   

\begin{figure}[t]
\centering
\includegraphics[width=.98\linewidth]{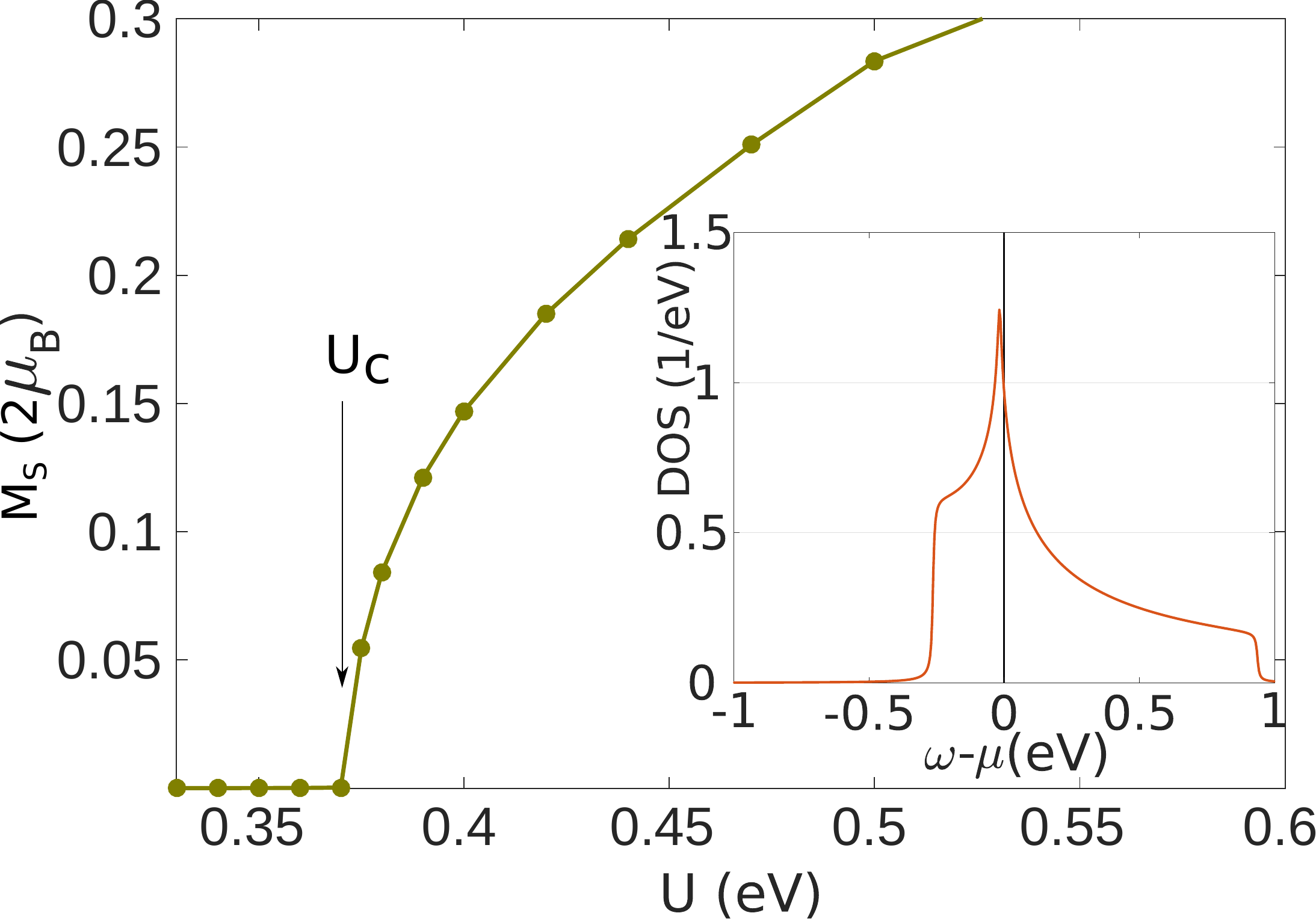}
\caption{Average staggered  magnetization $M_s$ 
in the homogeneous monolayer $30\times30$ system as a function of Hubbard repulsion $U$ at a filling of $n=0.80$ and temperature $kT=0.015\ \mathrm{eV}$.
 The orange solid curve in the inset is the DOS for the homogeneous monolayer with   hoppings as discussed in the text, at filling $n=0.80$.
}
\label{fig:DOS_and_UvsM}
\end{figure}

 The tight-binding density of states  for the homogeneous monolayer system, important in determining the ultimate magnetic response of the system we construct, is shown in Fig. \ref{fig:DOS_and_UvsM}.
 The critical Hubbard repulsion $U_c$ for long-range AFM order in the homogeneous lattice in absence of any dislocation or/and charge build up is located at 0.37 eV for this particular choice of parameters, as seen in the figure.
 
Note that the choice of interactions and filling of the model considered here does not correspond to any specific cuprate.  While disorder-induced magnetism has been observed, e.g. in LSCO up to approximately 15\% doping, of the same order as the doping chosen here, the exact values of these parameters are not particularly important.
Our aim is simply to generate a magnetic state by choosing $U$ sufficiently close to the critical $U$ for homogeneous magnetism as shown in Fig. \ref{fig:DOS_and_UvsM}.  Note further that the quantity $U$ in mean field theory, which generally overestimates magnetic order, is an effective parameter\cite{Bulut1993}, not to be compared to bare $U$ values from Hubbard model simulations.

 \subsection{Local dislocation-induced magnetic states}
 We expect that local defect-induced magnetic states, if stable, will be nucleated for correlations  slightly less than this value.   Note that the position of the chemical potential above the van Hove singularity is important within the current model, as the buildup of excess charge  around the defect reduces the chemical potential locally and therefore drives the system towards the van Hove singularity and  through the Stoner instability.  We believe that a  treatment of the disorder-induced magnetic effects beyond mean field would not require this fine tuning, but the current approach is simple, transparent, and one of the few methods applicable to treat inhomogeneous systems.   
\begin{figure*}[htpb]
\centering
\includegraphics[width=1\textwidth]{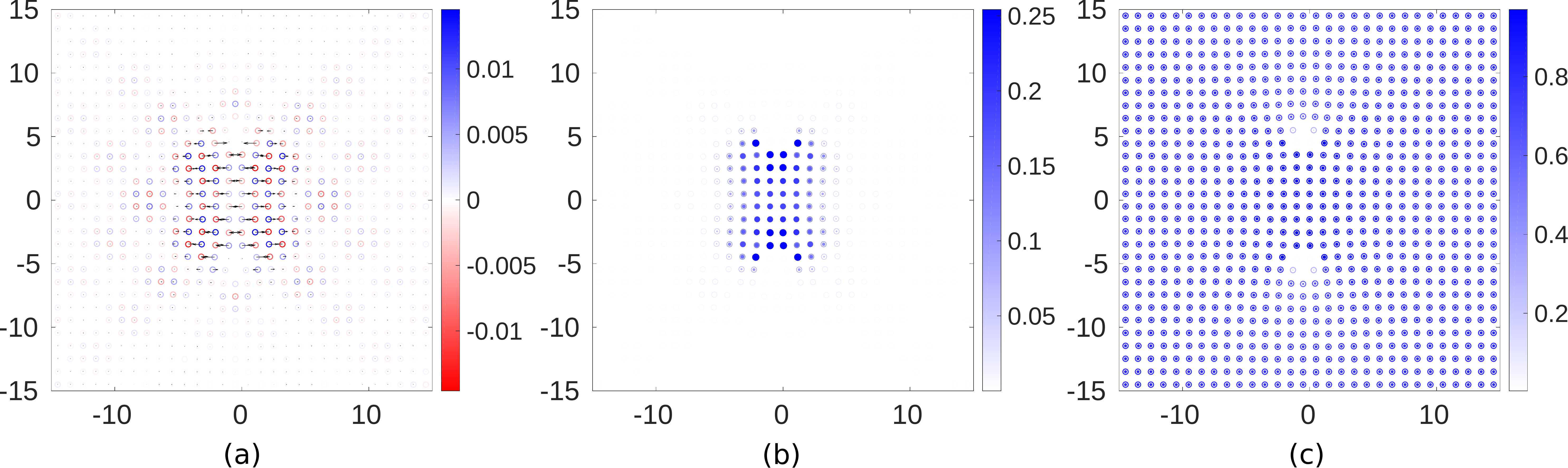}
\caption{Dislocation induces a local magnetization for a repulsion strength just below $U_c$ with a system energy of $E=-2.0403673\times 10^{-1}\ \mathrm{eV}$: Layer $\nu=1$, $kT=0.015\ \mathrm{eV},\ U=0.34\ \mathrm{eV},\ n=0.8,\ N=30,\ \mathbf{h}=0$.
(a) Magnetization (the arrows and the color scale show magnetization parallel  and normal to the lattice respectively),\ $|\textbf{M}|_{xy,max}=0.25426$
    ,\ Avg. $\textbf{M}=(-5.6119\times10^{-8},\ -1.4211\times10^{-6},\ 9.6084\times10^{-9})$, (b) Magnitude of magnetization, $|\textbf{M}|_{max}=0.25429$, (c) density. }
    \label{fig:layer1_no_field}
\end{figure*}
\begin{figure*}[htpb]
\centering
\includegraphics[width=1\textwidth]{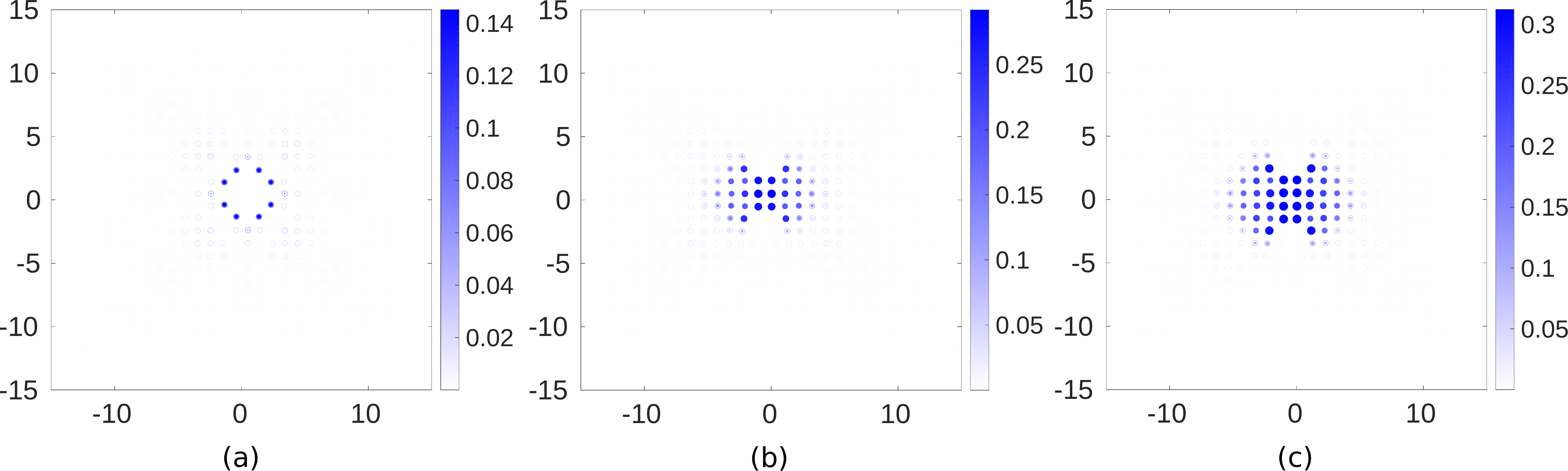}
\caption{Evolution of the magnitude of the  magnetization  as dislocation length increases (layer $\nu=1$): $kT=0.015\ \mathrm{eV},\ U=0.34\ \mathrm{eV},\ n=0.8,\ N=30$ (a) point dislocation (b) $5$ site dislocation (c) $6$ site dislocation. The case of a point dislocation (removed site, with accompanying relaxation of the lattice and charging) is  $C_4$ symmetric.}
\label{fig:varying_disloc_length}
\end{figure*}

Choosing a $U$ that is below but sufficiently close ($0.34\ \mathrm{eV}$) to $U_c$, we  can study  local AFM order both in  monolayer and bilayer distorted lattices. Fig. \ref{fig:layer1_no_field} shows the configuration of magnetization direction, magnitude, and electron charge for one layer of a bilayer lattice in presence of both interlayer hybridization and spin-orbit coupling for a dislocation of ten sites.   In our model including spin-orbit coupling, the second layer is related to the first by reversing the direction of the effective (in-plane) magnetic field, thus without external field, the in-plane magnetization of the converged systems is also just reversed and therefore not presented here. 
In Fig. \ref{fig:layer1_no_field}, one can see that although the magnetization is highly localized, there is some weak oscillating large distance component of magnetization normal to the plane. This may possibly reflect some finite size effect, due to the interference of the defect state with its periodically repeated copies.  One can  eliminate this effect by working at significantly larger lattices;   however, in a real system with a high concentration of defects with an inter-defect distance comparable to the system size shown, such interference will certainly exist. 

 It is interesting to examine how the magnetic state induced by a localized nonmagnetic perturbation here differs from the usual picture of an impurity-induced magnetic ``puddle" or  ``island''\cite{4mousq}.  As shown in Fig. \ref{fig:varying_disloc_length}(a), already for a single site defect, the response in our model differs substantially from the response of the metallic system to  a strong impurity placed on the regular lattice.  In the absence of a gap in the system, a single strong impurity does not produce a bound state due to its coupling to the metallic continuum, and this renders the formation of magnetic islands quite unlikely.  Clusters of strong impurities on regular lattices are known to create localized defect states and magnetism, however\cite{Schmid2010}.  In our model,  the lattice relaxes around the central missing site, and creates an extended one-body perturbation of the homogeneous Hamiltonian in both hoppings and onsite potentials.  This is analogous to the cluster of potentials on the regular lattice, and so indeed a magnetic state is created. Fig. \ref{fig:varying_disloc_length}(b) and (c) show further how this magnetic state evolves as sites are removed along a line.   It is important to note that the range of the magnetic impurity state is several times larger than the length of the dislocation itself.

\begin{figure*}[htpb]
\centering
\includegraphics[width=1\textwidth]{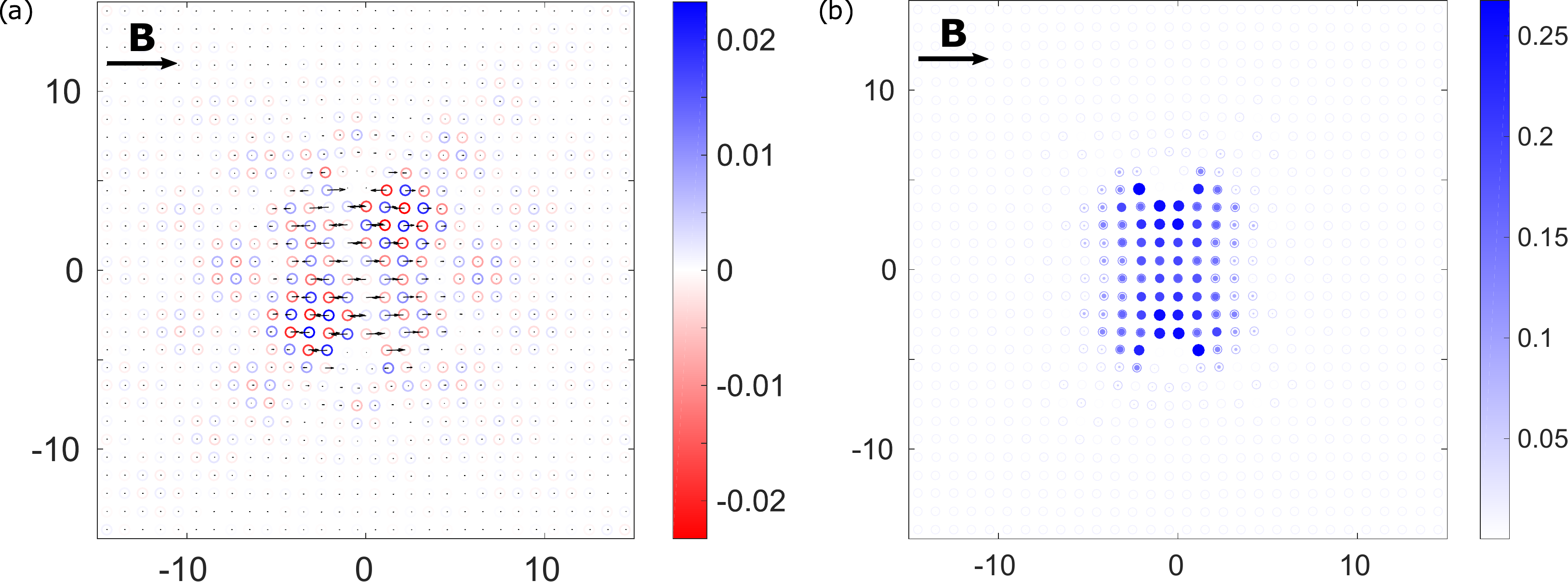}
\caption{Perturbing the locally magnetic configuration with a magnetic field along x axis,  leads to a system energy of $E_{h_x}=-2.0404780\times10^{-1}\ \mathrm{eV}$: Layer $\nu=1$, $kT=0.015\ \mathrm{eV},\ U=0.34\ \mathrm{eV},\ n=0.8,\ N=30,\ \mathbf{h}=(0.002,\ 0,\ 0)\ \mathrm{eV}$.
(a) Magnetization (the arrows and the color scale show magnetization parallel  and normal to the lattice respectively),\ $|\textbf{M}|_{xy,\mathrm{max}}=0.26754$,\ Avg. $\textbf{M}=(0.015133,\ -8.6289\times10^{-6},\ 1.3013\times10^{-8})$,\ (b) Magnitude of magnetization, $|\textbf{M}|_\mathrm{max}=0.26763$. A slight  enhancement of the magnetic order normal to the plane is visible in the top right and bottom left part of the magnetic region in (a). Magnetic order parallel to the plane enhances slightly at the top left and bottom right part,  also reflected in the  magnetization magnitude in (b).}   
    \label{fig:layer1_Bx}    
\end{figure*}

\begin{figure*}[htpb]
\centering
\includegraphics[width=1\textwidth]{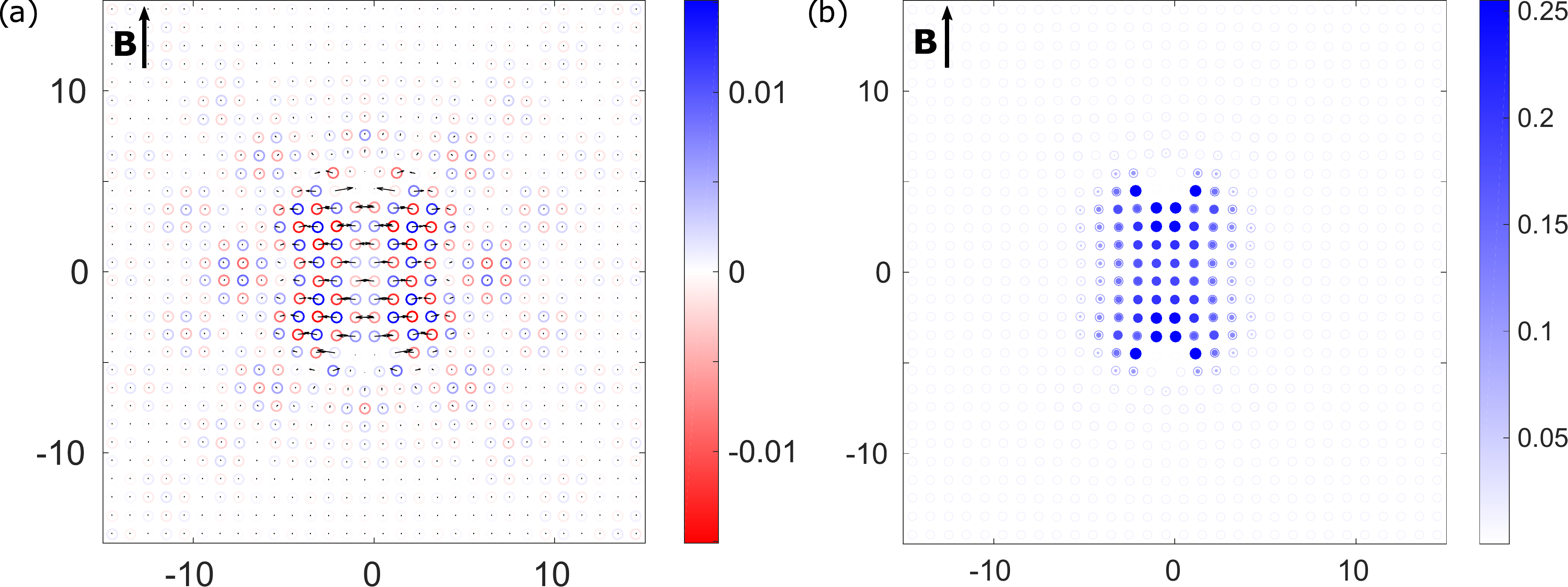}
    \caption{Perturbing the locally magnetic configuration with a magnetic field along y axis,  leading to a system energy of $E_{h_y}=-2.0405118\times10^{-1}\ \mathrm{eV}$: layer $\nu=1$, $kT=0.015\ \mathrm{eV},\ U=0.34\ \mathrm{eV},\ n=0.8,\ N=30,\ \textbf{h}=(0,\ 0.002,\ 0)\ \mathrm{eV}$.\
    (a) Magnetization, $|\textbf{M}|_{xy,\mathrm{max}}=0.25505$, Avg. $\textbf{M}=(-3.3947\times10^{-6},\ 0.015442,\ 1.4112\times10^{-8})$, (b) Magnitude of magnetization, $|\textbf{M}|_\mathrm{max}=0.25505$. 
    } 
    \label{fig:layer1_By}   
\end{figure*}
\subsection{Response to Zeeman field}
\par Even if the magnetic structures shown in Figs. \ref{fig:layer1_no_field} and \ref{fig:varying_disloc_length} break $C_4$ symmetry in the case of dislocation of length $>1$,  the system exhibits anisotropy in the  response to an external magnetic field only if the spin-orbit interaction is present. The magnetic response and corresponding energy on application of an in-plane magnetic field both normal and parallel to the line of dislocation are shown in Figs. \ref{fig:layer1_Bx} and Fig \ref{fig:layer1_By}. In the figures we show only the configuration of layer $\nu=1$, but they are based upon calculations for a bilayer; we give the energies for the entire  system.  The layer $\nu=0$ exhibits again a magnetization that would correspond to that of  layer $\nu=1$ if additionally the in-plane direction of the external field was flipped.  The local magnetic structure  without any field has inversion symmetry and also mirror symmetry w.r.t. both x and y axes within the layer as far as the magnitude of the  magnetization is concerned. For a magnetic  field applied normal to the line of dislocation, one can see that the mirror symmetry is lifted slightly but the inversion symmetry still persists.
The small but observable difference in total energies in presence of same magnetic field in $x$ and $y$ directions  will result in a detectable torque in magnetometry experiments. 

\section{Comparison with Experiments}
 Our goal here is to estimate whether magnetic structures created by such defects, e.g. in Fe-based superconductors, can for a reasonable defect density create an observable effect in torque magnetometry experiments.
To estimate the torque, we consider the energy difference $\Delta E=E_{h_x}-E_{h_y}$ of the systems converged in presence of magnetic fields applied along x and y directions and divide this by the rotation angle of $\pi /2$ to get an average torque for one dislocation in our system of  $30\times30$ lattice points.
If a $30\times30$ lattice, which spans $\sim12\ \mathrm{nm}\times 12\ \mathrm{nm}$, contains one dislocation, as we assumed in our simulation, a thin film of size { $\sim 100\ \mathrm{\mu m}\times 100\ \mathrm{\mu m}$} which is in the typical range \cite{PhysRevLett.103.157003}  for thin film torque magnetometry, would have $\sim10^8$ dislocations that would crudely give rise to more than $2.2 \times 10^{-17}$ N\,m/rad of torque per bi-layer, i.e. $\sim2.2 \times 10^{-13}$ N\,m/rad in the entire sample well within the measurable 
range \cite{Modic2014}.   In most of the cuprate phase diagram, however, the conditions to realize the magnetic state may not be present. Our calculations may apply best to bilayer systems like YBCO or BSCCO in the so-called spin glass state between the Mott insulator and the onset of superconductivity, where $\mu$SR experiments have reported signals of considerable magnetic disorder.

One dislocation per 100 nm$^2$ is also roughly the concentration of the linear defects visible in conductance maps of NaFeAs\cite{Rosenthal2014} above the tetragonal-orthorhombic transition $T_s$. 
This is an unusually large density of dislocations in a simple metal, but may be more generally realistic in some Fe-based systems, where strong magnetoelastic couplings are known to exist\cite{Boehmer2016}. If we assume similar defects are present in BaFe$_2$(As$_{1-x}$P$_x$)$_2$ in similar concentrations,  and account also for the fact that spin-orbit energies are an order of magnitude larger in FeSC than in cuprates, it is clear that a single domain sample of $\sim$100 $\mu$m $\times$ 100 $\mu$m  
with strain-induced dislocations may indeed account for the measurements.  Although these estimates are crude, our main goal in this work was to check the plausibility of the strain scenario, and it appears as though it cannot be ruled out.  More measurements and visualization methods  of the defects present are needed to clarify this explanation in the systems where nematic behavior above $T_s$ has been reported.   

Let us now estimate the torque to be expected on an underdoped cuprate sample with in-plane lattice constant $\sim0.4\ \mathrm{nm}$ and typical dimensions $250\times50\ \mathrm{\mu m}^3$ \cite{Matsuda2017}.  For example, a chain in a  YBa$_2$Cu$_3$O$_{7-x}$ sample with $x=0.7$  is expected to have a cumulative length of $\sim1.87\times10^5$ lattice constants of remnant single Cu-O chain  (assuming all the oxygen removal happens from Cu-O chain during doping). NMR studies suggest a minimum chainlet defect length of $50$ lattice constants \cite{YAMANI2004227} leading to an estimated $\sim7\times10^{8}$ number of chainlet defects in a single layer of the sample. Since within a unit cell of $c$-axis dimension 11.68\AA{} there are two chain layers, a sample of thickness $50\ \mathrm{\mu m}$ having $\sim8.56\times10^4$ layers would therefore contain $\sim6\times10^{13}$ chainlet defects. If a single chainlet defect is assumed to give rise to similar torque as a single dislocation as found in our calculation (i.e. $\sim2\times10^{-6}\mathrm{ eV}\approx3.2\times10^{-25}\ \mathrm{Nm}$), the entire sample film should experience a torque of $\sim1.19\times10^{-11}\ \mathrm{Nm}$ which agrees roughly with the observed torque response range \cite{Matsuda2017}. The susceptibility anisotropy $\eta=\frac{M_{yy}-M_{xx}}{M_{yy}+M_{xx}}\approx1.011\times10^{-2}$ (where $M_{xx}$, $M_{yy}$ are magnetization normal and parallel to the line defect (see Fig. \ref{fig:layer1_Bx}, \ref{fig:layer1_By}) for same perturbing field strength) although slightly large, is also in agreement in order of magnitude. Thus, we expect magnetic chainlet vacancy defects for $x$ less than $\lesssim 0.5$ to produce a torque of magnitude equal to or larger than the the observed torque signal at higher doping despite the near depletion of one of the chains.

\section{Conclusions}
In summary, we have discussed how
linear defects in correlated electron systems can create $C_2$ symmetric localized magnetic states, which can then couple to an external field via spin-orbit interaction.  We presented concrete calculations for a bilayer model of the cuprates appropriate for YBCO-123, where oxygen vacancy chainlets are known to produce such magnetic states\cite{Chen2009}, and estimated the torque magnetometry signal to be observed in experiments.  First we simulated a dislocation-antidislocation pair in a periodic lattice  with molecular dynamics. Next, we determined the critical Hubbard repulsion $U_c$ in the corresponding homogeneous lattice with suitable filling in absence of any kind of dislocation.   For correlation strengths $U\lesssim U_c$, localized magnetic dislocation states with symmetry lower than $C_4$ are nucleated and couple via spin-orbit coupling to an external magnetic field.  These states are rather interesting in their own right, and may be detected on systems with atomically smooth surfaces by spin-polarized STM.  Our calculations are crudely consistent with existing torque magnetometry signals detected in YBCO\cite{Matsuda2017}; moreover, we predict enhanced nematic signals as doping
is lowered further beyond those dopings studied in experiment, to O6.3-6.4 concentrations where strong nematic signals have been detected in neutron scattering\cite{Keimer2008} and transport\cite{Ando2002}.
 
We further believe our calculations are also highly relevant for issues of nematicity that arise in the Fe-based superconductivity field.    In particular, we have shown that built-in strain  in a two dimensional lattice can create localized magnetic dislocation
states with symmetry lower than that of the surrounding lattice, and thereby give rise to signals of nematic behavior in the system even if the system is in a nominal tetragonal phase.  Our estimates of the torque magnetometry signal arising from such defects confirm the earlier suggestion\cite{Rosenthal2014} of strain 
as a possible explanation for nematic signals observed above the tetragonal-orthorhombic transition in Fe-based superconductors, in the presence of spin-orbit interaction, and place it on a concrete foundation.  While we have not yet performed realistic multiorbital calculations, our one-band calculations will be straightforward to generalize. The torque on a sample containing a single domain of aligned dislocations, such as apparently observed by STM in the Fe-based superconductor NaFeAs, was estimated from the one-band result and found to be easily detectable by  current torque magnetometry techniques.

\vspace{0.2cm}

\textit{Acknowledgements:} We thank H. Christiansen, R. Fernandes, R. Hennig, F. M\"uller, A. Nevidomskyy and T. Shibauchi for useful discussions. This work was supported in part by  
NSF-DMR-1849751. 

\bibliography{dislocation}
\end{document}